\documentclass[aps,prb,twocolumn,showpacs]{revtex4}

\usepackage{graphicx}

\begin{document}

\title{Berry phase of magnons in textured ferromagnets}
\author{V.~K.~Dugaev$^{1,2,\dag }$, P. Bruno$^1$, B. Canals$^3$, and C. Lacroix$^3$}
\affiliation{$^1$Max-Planck-Institut f\"ur Mikrostrukturphysik,
Weinberg 2, 06120 Halle, Germany\\
$^2$Department of Physics and CFIF, Instituto Superior
T\'ecnico, Av. Rovisco Pais, 1049-001 Lisbon, Portugal\\
$^3$Laboratoire Louis N\'eel, CNRS, BP 166, 38042 Grenoble,
Cedex 09, France}
\date{\today }

\begin{abstract}
We study the energy spectrum of magnons in a ferromagnet with
topologically nontrivial magnetization profile. In the case of
inhomogeneous magnetization corresponding to a metastable state of
ferromagnet, the spin-wave equation of motion acquires a gauge
potential leading to a Berry phase for the magnons propagating
along a closed contour. The effect of magnetic anisotropy is
crucial for the Berry phase: we show that the anisotropy
suppresses its magnitude, which makes the Berry phase observable
in some cases, similar to the Aharonov-Bohm effect for electrons.
For example, it can be observed in the interference of spin waves
propagating in mesoscopic rings. We discuss the effect of domain
walls on the interference in ferromagnetic rings, and propose some
experiments with a certain geometry of magnetization. We also show 
that the nonvanishing average topological field acts on the magnons
like a uniform magnetic field on electrons. It leads to the
quantization of the magnon spectrum in the topological field.

\vskip0.5cm \noindent
\end{abstract}
\maketitle

\section{Introduction}

The Berry phase theory \cite{berry84,shapere89,bohm03} allowed to
generalize the idea of Aharonov-Bohm effect \cite{aharonov59} on
electrons in the electromagnetic potential, to an analogous effect
related to a gauge potential, which arises during the adiabatic
motion of a quantum system in a parametric space. Up to now a lot
of efforts has been directed to understand better and to find an
experimental confirmation for the Berry phase of electrons like,
for example, in the case of electrons moving in a varying
magnetization field of the inhomogeneous ferromagnet.
\cite{lyanda98,ye99}

One of the most intriguing consequences of the Berry phase theory
is a possibility of the Aharonov-Bohm-like effect on electrically
neutral particles or boson fields.\cite{anandan82,chiao86} An
example of the adiabatic phase for the polarized light has been
investigated by Pancharatnam \cite{pancharatnam56} and Berry.
\cite{berry87} The other example is the Aharonov-Bohm effect for
the exciton,\cite{romer00} which is a bound state of an electron
and a hole in semiconductors.

Here we consider the effect of the gauge potential and Berry phase
on the propagation of magnons in textured ferromagnets. Such
quasiparticles are usually viewed as the elementary excitations of
the ordered homogeneous state of a ferromagnet but they can be
also used to classify the excited states near a metastable
inhomogeneous magnetic configuration. These magnons describe the
dynamics of weakly excited inhomogeneous ferromagnet.

The dynamics of magnetization in nanomagnets is in the focus of
recent activity\cite{dynamics} because of the importance of this
problem for magnetoelectronic applications.\cite{wolf01,zutic04}
It includes the switching of magnetization by electric current,
spin pumping, magnetization reversal in microscopic spin valves,
etc. Usually, the magnons play a negative role in the
magnetization dynamics limiting the frequency of magnetic
reversal, and also leading to the energy dissipation. However,
they can be probably used in the spin transport phenomena like the
spin currents of magnetically polarized electrons.

Here we study the energy spectrum of spin waves in ferromagnets
with a static inhomogeneous magnetization profile, and we
demonstrate a possibility of observation of the Berry phase in
the interference experiments on spin waves in magnetic
nanostructures. Recent results of the micromagnetic computer
simulation\cite{yamasaki03,ha03,hertel03} of such systems
demonstrate that the interference of spin waves can be really
observed in magnetic nanorings with domain walls.

The equation for spin wave excitations in a general case of
arbitrary local frame depending on both coordinate and time, has
been found long ago by Korenman {\it et al}\cite{korenman77} in
the context of local-band theory of itinerant
magnetism.\cite{korenman} Here we use an idea of this method to relate
the adiabatic space transformation to the Berry phase and to find
corresponding properties of the spin waves in a topologically
nontrivial inhomogeneous magnetic profile,
which is a metastable state of the ferromagnet. We
show that the magnetic anisotropy is a crucial element determining
the possibility of observation of the Berry phase in real
experiments.

\section{Model and spin wave equations}

We consider the model of a ferromagnet described by the Hamiltonian,
which includes the exchange interaction, anisotropy, and the
interaction with external magnetic field . It
has the following general form
\begin{eqnarray}
\label{1}
H=\frac12 \int d^3{\bf r}\left[
a\left( \frac{\partial n_\mu ({\bf r})}{\partial r_i}\right ) ^2
+\mathcal{F}\left\{ {\bf n}({\bf r})\right\}
\right] ,
\end{eqnarray}
where ${\bf n}({\bf r})$ is the unit vector oriented along
the magnetization ${\bf M}({\bf r})$ at the point ${\bf r}$,
$a$ is the constant
of exchange interaction, $\mathcal{F}\left\{ {\bf n}({\bf r})\right\}$
is a function determining the magnetic anisotropy
[correspondingly, it includes a certain number
of tensors relating the components of vector ${\bf n}({\bf r})$]
and the dependence on external field,
and $M_0$ is the magnitude of magnetization.

Due to the condition ${\bf n}^2({\bf r})=1$, the model is
constrained and belongs to the class of nonlinear $\sigma $
models.\cite{nagaosa99}
The stationary (saddle point) solutions for the magnetization
vector ${\bf n}_0({\bf r})$
describing metastable states of the ferromagnet, can be found
by minimizing Hamiltonian (1)
with the constraint ${\bf n}^2({\bf r})=1$.
It was shown (see, e.g., Refs.~[\onlinecite{belavin75,pokrovsky88}])
that such metastable states with inhomogeneous magnetization profile
are related to the topology of ferromagnetic ordering, and they can
include skyrmions, magnetic vortices, and other topological objects.

We will be interested in describing the dynamics of small
deviations $\delta {\bf n}({\bf r})$ from a certain
metastable profile ${\bf n}_0({\bf r})$
with a nonuniform magnetization,
${\bf n}({\bf r})={\bf n}_0({\bf r})+\delta {\bf n}({\bf r})$,
$\left| \delta {\bf n}({\bf r})\right| \ll 1$.
Correspondingly,
we assume that the solution of a saddle-point equation
describing the state ${\bf n}_0({\bf r})$ is
already known.

We perform a local transformation
\begin{equation}
\label{2}
\tilde{\bf n}({\bf r})={\rm R}({\bf r})\, {\bf n}({\bf r}),
\end{equation}
using the orthogonal transformation matrix
${\rm R}({\bf r})$. By definition, it determines the rotation of
local frame in each point of the space,
so that the magnetization in the local frame is oriented
along the $z$ axis, $\tilde{\bf n}_0=(0,\, 0,\, 1)$.
Then we consider small deviations of magnetization ${\bf s}({\bf r})$
from $\tilde{\bf n}_0$. Since ${\bf s}({\bf r})$ is small and vectors
$\tilde{\bf n}_0$ are oriented along $z$,
the vectors ${\bf s}({\bf r})$ lie in the $x$-$y$ plane.

The transformation matrix in Eq.~(2) is taken in a general form of
orthogonal transformation
\begin{equation}
\label{3}
{\rm R}({\bf r})
=e^{i\psi ({\bf r})\, J_z}\,
e^{i\theta ({\bf r})\, J_y}\,
e^{i\phi ({\bf r})\, J_z}\, ,
\end{equation}
where $\psi $, $\theta $, $\phi $ are the
Euler angles determining an arbitrary rotation of the
coordinate frame,
and $J_x$, $J_y$, and $J_z$ are the generators of
3D rotations around $x$, $y$ and $z$ axes.

Two rotation parameters (for definiteness, the angles $\theta $
and $\phi $) can be used to define the frame with the $z$ axis along
the vector ${\bf n}_0({\bf r})$. In the absence of anisotropy,
the additional rotation to the angle $\psi $ is purely gauge transformation.
However, in a general case of anisotropic system, this rotation
allows to choose the local frame in correspondence with the
orientation of anisotropy axes.

The Hamiltonian of exchange interaction (the first term in Eq.~(1))
in the rotated frame has the following form
\begin{equation}
\label{4}
H_{ex}=\frac{a}2 \int d^3{\bf r}
\left( \frac{\partial \tilde{n}_\mu }{\partial r_i}
-A_i^{\mu\nu}\tilde{n}_\nu \right) ^2,
\end{equation}
where the gauge field $A_i({\bf r})$ is defined by
\begin{equation}
\label{5}
A_i({\bf r})=\left( \frac{\partial }{\partial r_i}\; {\rm R}\right) \, {\rm R}^{-1}.
\end{equation}
Transformation (3) and gauge potential (5) are $3\times 3$ matrices
acting on the magnetization vectors. The matrix $A_i({\bf r})$ can be
also presented as
\begin{equation}
\label{6}
A_i({\bf r})=i\mathcal{A}^\mu _i({\bf r})\, J_\mu ,
\end{equation}
where $\mathcal{A}^\mu _i({\bf r})$ belongs to the adjoint representation
of the rotation group.

Using (3) and (6), we find the explicit dependence of the gauge potential
on the Euler angles
\begin{eqnarray}
\label{7}
\mathcal{A}^x_i({\bf r})
=\sin \psi \; \frac{\partial \theta }{\partial r_i}
-\sin \theta \; \cos \psi \; \frac{\partial \phi }{\partial r_i}\; ,
\nonumber \\
\mathcal{A}^y_i({\bf r})
=\cos \psi \; \frac{\partial \theta }{\partial r_i}
+\sin \theta \sin \psi \; \frac{\partial \phi }{\partial r_i}\; ,
\\
\mathcal{A}^z_i({\bf r})
=\frac{\partial \psi }{\partial r_i}
+\cos \theta \; \frac{\partial \phi }{\partial r_i}\; .\hskip1cm
\nonumber
\end{eqnarray}

The magnetic anisotropy described by the second term in the right
hand part of (1) gives after transformation to the local
frame a function
$\tilde{\mathcal{F}}\left\{ \tilde{\bf n}({\bf r})\right\} $
with correspondingly transformed tensor fields. Here we do not
restrict the general consideration of the problem by any
specific form of the anisotropy but in the following
we consider the most important examples of easy
plane and easy axis anisotropy.

The Landau-Lifshitz equations for the magnetization in the locally
transformed frame are
\begin{equation}
\label{8}
\frac{\partial \tilde{n}_\mu }{\partial t}
=-\frac{\gamma }{M_0}\; \epsilon _{\mu\nu\lambda}\, \tilde{n}_\nu
\left( \frac{\delta H}{\delta \tilde{n}_\lambda }
-\nabla ^{\lambda\rho}_i\,
\frac{\delta H}{\delta \, (\nabla ^{\rho\tau}_i\, \tilde{n}_\tau )}
\right) ,
\end{equation}
where $\epsilon _{\mu\nu\lambda}$ is the unit antisymmetric tensor, and
\begin{equation}
\label{9}
\nabla ^{\mu\nu}_i
=\frac{\partial}{\partial r_i}\;
\delta _{\mu\nu}-A_i^{\mu\nu}
\end{equation}
is the covariant derivative. The right-hand part of Eq.~(8)
vanishes for the magnetization profile corresponding to a
metastable state. This is seen from the Landau-Lifshitz
equation in the unrotated original frame. In the following,
we will use Eq.~(8) for the small deviations of magnetization
from the metastable state. Hence, we will consider in the right
part of (8) only the terms linear in deviations.

Using (1), (4) and (8) we find the equations for weak magnetic
excitations near the metastable state (spin waves)
\begin{eqnarray}
\label{10}
\frac{\partial s_x}{\partial t} =-c_s\left[
\frac{\partial ^2s_y}{\partial r_i^2} -2\mathcal{A}^z_i\,
\frac{\partial s_x}{\partial r_i} -(\mathcal{A}^z_i)^2\, s_y
\right. \nonumber \\ \left.
-\frac{\partial \mathcal{A}^z_i}{\partial r_i}\; s_x
+\, (\mathcal{A}^y_i)^2\, s_y +\mathcal{A}^x_i\,
\mathcal{A}^y_i\, s_x \right]
\nonumber \\
+\frac{\gamma }{M_0}\; \frac{\partial ^2\tilde{\mathcal{F}}}
{\partial \tilde{n}_x\, \partial \tilde{n}_y}\; s_x
+\, \frac{\gamma }{M_0}\; \frac{\partial ^2\tilde{\mathcal{F}}}
{\partial \tilde{n}_y^2}\; s_y
\; ,
\end{eqnarray}
\begin{eqnarray}
\label{11}
\frac{\partial s_y}{\partial t}
=c_s\left[
\frac{\partial ^2s_x}{\partial r_i^2}
+2\mathcal{A}^z_i\, \frac{\partial s_y}{\partial r_i}
-(\mathcal{A}^z_i)^2s_x
\right. \nonumber \\ \left.
+\frac{\partial \mathcal{A}^z_i}{\partial r_i}\; s_y
+\, (\mathcal{A}^x_i)^2\, s_x
+\mathcal{A}^x_i\, \mathcal{A}^y_i\, s_y
\right]
\nonumber \\
-\frac{\gamma }{M_0}\; \frac{\partial ^2\tilde{\mathcal{F}}}
{\partial \tilde{n}_x\, \partial \tilde{n}_y}\; s_y
-\, \frac{\gamma }{M_0}\; \frac{\partial ^2\tilde{\mathcal{F}}}
{\partial \tilde{n}_x^2}\; s_x
\; ,
\end{eqnarray}
where $c_s=\gamma a/M_0$ is the stiffness.

Using (10) and (11)
we can also present the equations for circular
components of the spin wave, $s_\pm =s_x\pm is_y$,
\begin{eqnarray}
\label{12}
\pm \, i\; \frac{\partial s_\pm }{\partial t}
=\left[ -c_s\left( \frac{\partial }{\partial r_i}
\mp iA^z_i \right) ^2
-V({\bf r})
+\frac{\gamma }{2M_0}\; \frac{\partial ^2\tilde{\mathcal{F}}}
{\partial \tilde{n}_x^2}
\right. \nonumber \\ \left.
+\frac{\gamma }{2M_0}\; \frac{\partial ^2\tilde{\mathcal{F}}}
{\partial \tilde{n}_y^2}
\right]
s_\pm
+\left[
-w({\bf r})
-ic_s\, \mathcal{A}^x_i\mathcal{A}^y_i
\right. \nonumber \\ \left.
+\, \frac{i\gamma}{M_0}\; \frac{\partial ^2\tilde{\mathcal{F}}}
{\partial \tilde{n}_x\, \partial \tilde{n}_y}
+\frac{\gamma }{2M_0}\; \frac{\partial ^2\tilde{\mathcal{F}}}
{\partial \tilde{n}_x^2}
-\frac{\gamma }{2M_0}\; \frac{\partial ^2\tilde{\mathcal{F}}}
{\partial \tilde{n}_y^2}
\right] s_\mp \, ,\hskip0.3cm
\end{eqnarray}
where $V({\bf r})$ and $w({\bf r})$ are, respectively,
the effective potential and a mixing field acting on the spin wave:
\begin{equation}
\label{13}
V({\bf r})=\frac{c_s}2\, \left[
\left( \mathcal{A}^x_i\right) ^2+\left( \mathcal{A}^y_i\right) ^2
\right] ,
\end{equation}
\begin{equation}
\label{14}
w ({\bf r})=\frac{c_s}2\, \left[
\left( \mathcal{A}^x_i\right) ^2-\left( \mathcal{A}^y_i\right) ^2
\right] .
\end{equation}
Equations (12) for $s_+({\bf r},t)$ and $s_-({\bf r},t)$ are complex
conjugate to each other since they both describe the same spin
wave with real components $s_x({\bf r},t)$ and $s_y({\bf r},t)$.

We can see that $V({\bf r})$ is an effective potential profile
for the propagation of spin wave.
Due to the terms $w ({\bf r})$ and $ic_s\mathcal{A}_i^x\mathcal{A}_i^y$
in (12), the equations for circular
components $s_+$ and $s_-$ are coupled even in the absence of
anisotropy.
All these terms are of the second order in derivative of the rotation angle,
and they are small in the adiabatic limit corresponding
to a smooth variation of the magnetization vector ${\bf n}({\bf r})$.

\section{Semiclassical approximation}

Equations (10) and (11) can be solved in the semiclassical approximation.
The condition of its applicability is a smooth variation of gauge
potential $\mathcal{A}^\mu _i({\bf r})$ and fields related to the anisotropy,
as well as the external magnetic field, at the
wavelength of the spin wave,
$kL\gg 1$, where $k$ is the wavevector of the spin wave and $L$ is the
characteristic length of the variation of
$\mathcal{A}^\mu _i({\bf r})$ and $\mathcal{F}\left\{ {\bf n}({\bf r})\right\} $
(more exactly, the minimum of the corresponding characteristic lengths).
Note that the condition of applicability of the semiclassical
approximation to solve the spin-wave equations, does not require any
smallness of the gauge potential itself.

Starting from Eqs.~(10) and (11), we look for a general semiclassical
solution in the form
\begin{equation}
\label{15}
s_x({\bf r},t)=
a\, \cos \left[ \xi ({\bf r})-\omega t\right]
+b\, \sin \left[ \xi ({\bf r})-\omega t\right] ,
\end{equation}
\begin{equation}
\label{16}
s_y({\bf r},t)=
d\, \sin \left[ \xi ({\bf r})-\omega t\right]
+f\, \cos \left[ \xi ({\bf r})-\omega t\right] ,
\end{equation}
with arbitrary coefficients $a$, $b$, $d$, $f$, and a smooth function
$\xi ({\bf r})$, so that we can neglect the second derivative of
$\xi ({\bf r})$ over coordinate ${\bf r}$.
Substituting (15) and (16) in (10) and (11), we can find four equation for
the $a$, $b$, $d$, $f$ coefficients.

The solution (15), (16) describes the elliptic spin wave with an arbitrary
choice of the axes $x$ and $y$, and, generally, with a varying in space orientation
of the principal axes of the ellipse. We can simplify our consideration
by choosing the angle $\psi ({\bf r})$ at each point of the space in accordance
with the orientation of the principal axes.
The corresponding equation for $\psi ({\bf r})$ can be found from the
condition of $b=f=0$ in Eqs.~(15) and (16)
\begin{equation}
\label{17}
c_s\, \mathcal{A}^x_i\, \mathcal{A}^y_i
-\frac{\gamma }{M_0}\; \frac{\partial ^2\tilde{\mathcal{F}}}
{\partial \tilde{n}_x\, \partial \tilde{n}_y}=0.
\end{equation}

Using (17) and neglecting the terms with derivative of $\mathcal{A}_i^z $, which are
small in the semiclassical approximation, we write the spin-wave equations
(10) and (11) as
\begin{eqnarray}
\label{18}
\frac{\partial s_x}{\partial t}
=-c_s\left[
\frac{\partial ^2s_y}{\partial r_i^2}
-2\mathcal{A}^z_i\, \frac{\partial s_x}{\partial r_i}
-(A^z_i)^2\, s_y
+\, (\mathcal{A}^y_i)^2\, s_y
\right]
\nonumber \\
+\, \frac{\gamma }{M_0}\; \frac{\partial ^2\tilde{\mathcal{F}}}
{\partial \tilde{n}_y^2}\; s_y\; ,\hskip0.5cm
\end{eqnarray}
\begin{eqnarray}
\label{19}
\frac{\partial s_y}{\partial t}
=c_s\left[
\frac{\partial ^2s_x}{\partial r_i^2}
+2\mathcal{A}^z_i\, \frac{\partial s_y}{\partial r_i}
-(\mathcal{A}^z_i)^2s_x
+\, (\mathcal{A}^x_i)^2\, s_x
\right]
\nonumber \\
-\, \frac{\gamma }{M_0}\; \frac{\partial ^2\tilde{\mathcal{F}}}
{\partial \tilde{n}_x^2}\; s_x\; .\hskip0.5cm
\end{eqnarray}
Note that by fixing the angle $\psi $ in Eq.~(17), we are choosing the gauge,
which defines completely the potential $\mathcal{A}_i^\mu $. We do it
in spirit of the usual fixing gauge in the WKB approximation.

After substitution of (15) and (16) with $b=f=0$ into (18) and (19),
we come to the following equation for the momentum
$k_i({\bf r})\equiv \partial \xi ({\bf r})/\partial r_i$
\begin{eqnarray}
\label{20}
\left( \omega +2c_s\, \mathcal{A}^z_i\, k_i\right) ^2
-\left\{
c_s\left[ k_i^2+(\mathcal{A}^z_i)^2-(\mathcal{A}^y_i)^2\right]
\right. \nonumber \\ \left.
+2p_x\right\}
\left\{
c_s\left[ k_i^2+(\mathcal{A}^z_i)^2-(\mathcal{A}^x_i)^2\right]
+2p_y\right\} =0,\hskip0.1cm
\end{eqnarray}
where
\begin{eqnarray}
\label{21}
p_{x,y}({\bf r})=\frac{\gamma }{2M_0}\; \frac{\partial ^2\tilde{\mathcal{F}}}
{\partial \tilde{n}_{x,y}^2}\;
\end{eqnarray}
are the anisotropy parameters.

Equation (20) should be solved for $k_i({\bf r})$ as a function of
smooth inhomogeneous field
$\mathcal{A}_i^\mu ({\bf r})$.
This equation does not constraint the orientation of ${\bf k}({\bf r})$ but
determines the magnitude of vector ${\bf k}({\bf r})$ for each
direction in the momentum space.
Let us take vector ${\bf k}({\bf r})$ along an arbitrary direction,
defined by a unity vector ${\bf g}$. Then we can rewrite (20) as
\begin{eqnarray}
\label{22}
\left[ \omega +2c_sk\, g_i\mathcal{A}^z_i\right] ^2
-\left\{
c_s\left[ k_i^2+(\mathcal{A}^z_i)^2-(\mathcal{A}^x_i)^2\right]
+2p_x\right\}
\nonumber \\
\times \left\{
c_s\left[ k_i^2+(\mathcal{A}^z_i)^2-(\mathcal{A}^y_i)^2\right]
+2p_y\right\} =0,\hskip0.5cm
\end{eqnarray}
and we come to the fourth-order algebraic equation for $k({\bf r})$.
It can be solved numerically, and a resulting dependence of
$k_i({\bf r})$ on the gauge field in the integral
$\xi ({\bf r})=\int _Ck_i({\bf r})\, dr_i$ leads to
the Berry phase acquired by the spin wave propagating along the contour $C$.

We can find the solution of Eq.~(22) analytically in the limit of
weak gauge potential
$\left| \mathcal{A}^z_i\right| \ll k$, which corresponds to the adiabatic
variation of the magnetization direction ${\bf n}({\bf r})$ and also the
adiabatic rotation in space of the elliptic trajectory,
$\left |\partial \psi /\partial r_i\right| \ll k$.
Then in the first order of $\mathcal{A}_i^z$ we find
\begin{eqnarray}
\label{23}
k_i({\bf r})
\simeq \frac{g_i}{\sqrt{c_s}}\left[
\left( \omega ^2+p^2\right) ^{1/2}
-p\, \right] ^{1/2}
\nonumber \\
+\, g_i\, g_j\, \mathcal{A}^z_j({\bf r})\,
\left( 1+p^2/\omega ^2 \right) ^{-1/2},
\end{eqnarray}
where $p=\left| p_x-p_y\right| $.

Using Eq.~(23) and taking the vector ${\bf g}$ along the tangent at
each point of a closed contour $C$, we find the Berry phase
\begin{equation}
\label{24}
\gamma _B(C)
=\oint _C\;
\frac{\mathcal{A}^z_i({\bf r})\, dr_i}
{\left( 1+p^2/\omega ^2 \right) ^{1/2}}\; .
\end{equation}
As follows from (24), the Berry phase $\gamma _B(C)$ in the
anisotropic system acquires an additional factor
$\kappa \equiv \left( 1+p^2/\omega ^2 \right) ^{-1/2}$ depending
on the magnetic anisotropy parameter $p$.

The denominator in (24) has a simple geometrical interpretation.
Indeed, the coefficients $a$ and $d$ in the semiclassical solution
(15), (16) are the ellipse parameters, which are
related to the anisotropy factor $p$
\begin{equation}
\label{25}
\frac{a}{d}
=\left( 1+\frac{p^2}{\omega ^2}\right) ^{1/2}
-\frac{p}{\omega }\; .
\end{equation}
Correspondingly, we can relate the parameter
$\kappa =\left( 1+p^2/\omega ^2 \right) ^{-1/2}$ in
Eq.~(24) to the geometry parameters of the ellipse
\begin{equation}
\label{26}
\kappa =\sin 2\Theta ,
\end{equation}
where $\Theta =\arctan (d/a)$.

Using definition (7), the Berry phase can be finally presented as
\begin{equation}
\label{27}
\gamma _B(C)
=\oint _C\;
\left[ \left( \kappa -1\right)
\frac{\partial (\psi +\phi )}{\partial r_i}
+\kappa \left( \cos \theta -1\right) \;
\frac{\partial \phi }{\partial r_i}\right] dr_i.
\end{equation}
In this expression we extracted a term proportional
to $2\pi N$, $N\in \mathbf{Z}$. This allows to avoid the
multivaluedness of Berry phase in the absence of
anisotropy when $\kappa =1$.\cite{bruno04}
The first term in (27) is proportional to the total winding number
of rotations associated with the angles $\psi $ and $\phi $, whereas
the second term is a spherical angle on $S_2$, which is the mapping space
of the vector field ${\bf n}({\bf r})$. The second term in (27) has
a standard interpretation of the Berry phase as the magnetic flux
penetrating
the contour on $S_2$, when the field is created by monopole at the center of
Berry sphere. Following this idea, one can interpret the first
term in (27) as the flux created by the magnetic string along the $z$ axis,
penetrating through the mapping contour on the unit circle.\cite{bruno04}
In accordance with Eq.~(27), this contribution to the Berry phase
vanishes for isotropic magnetic systems, $\kappa =1$.
The first term in (27) is the {\it topological} Berry phase
(it depends only on the winding number) in contrast to
the {\it geometric} Berry phase of the second term in (27).\cite{bruno04}

As follows from (24), the effective gauge field for spin waves in
the anisotropic
system is $\tilde{\mathcal{A}}_i=\kappa \mathcal{A}^z_i$, and
the corresponding topological field acting on the magnons
can be calculated as the curvature
of connection $\tilde{\mathcal A}_i({\bf r})$
\begin{equation}
\label{28}
B_i=\kappa \;
\epsilon _{ijk}\; \frac{\partial \mathcal{A}_k^z}{\partial r_j}
+\epsilon _{ijk}\; \frac{\partial \kappa }{\partial r_j}\; \mathcal{A}_k^z\, .
\end{equation}
Note that there is a contribution related to the variation in space of
the anisotropy parameters [second term in Eq.~(28)].

We consider now in more details the motion of elliptic spin wave in the
adiabatic regime. The anisotropy suppresses one of the components
$s_x$ or $s_y$ breaking the symmetry with respect to rotations
around $z$ axis.
Correspondingly, there is no gauge invariance
$s_+\rightarrow e^{i\varphi }s_+$ and
$s_-\rightarrow e^{-i\varphi }s_-$ for the circular components, and
the motion of magnetization in the spin wave is
elliptical. In the adiabatic limit of $\left| \mathcal{A}^z_i\right| \ll k$,
the solutions for
$s_x$ and $s_y$ are given by Eqs.~ (15) and (16) with $b=f=0$ and the
ratio of amplitudes $(a/d)$.
Thus, we could expect
the local invariance to transformations preserving the value
of $s_x^2+(d/a)^2s_y^2=const$ instead of simple rotations in $x-y$ plane.

Using the Fourier transformation
of Eqs.~(18) and (19) for $s_{x,y}$ we find the following equation
for the elliptic components of spin wave,
$\tilde{s}_\pm =s_x\pm i(d/a)\, s_y$
\begin{eqnarray}
\label{29}
\left[ \omega +2c_sk_i\mathcal{A}^z_i
-\tilde{c}_s\left\{ k_i^2+(\mathcal{A}^z_i)^2\right\}
-\frac{pd}{2a}\right] \tilde{s}_+
\nonumber \\
-\frac{a}{2d}
\left[ c_s\left\{ (k_i^2+(\mathcal{A}^z_i)^2\right\}\left( 1-\frac{d^2}{a^2}\right)
-\frac{pd^2}{a^2}\right] \tilde{s}_-=0,\hskip0.3cm
\end{eqnarray}
and the complex conjugate to (29),
where $\tilde{c}_s=c_s\, (1+d^2/a^2)a/2d\, $, and
we determine the $d/a$ from the condition of
vanishing of the second bracket in Eq.~(29). This condition determines
the ellipticity factor, and
we find that it coincides
with Eq.~(25) in the limit of $\left| \mathcal{A}^z_i\right| \ll k$.
Thus, we come to the following equation for the elliptic wave
in the gauge field
\begin{eqnarray}
\label{30}
\left\{ \omega +2c_sk_i\mathcal{A}^z_i
-\tilde{c}_s \left[ k_i^2+(\mathcal{A}^z_i)^2\right]
-\frac{pd}{2a}\right\} \tilde{s}_+=0.\hskip0.3cm
\end{eqnarray}
This equation is not gauge invariant but in the adiabatic regime, neglecting
the difference in small terms of the order of $(\mathcal{A}^z_i)^2$,
we can present it as
\begin{eqnarray}
\label{31}
\left[ \omega
-\tilde{c}_s \left( k_i-\kappa \, \mathcal{A}^z_i\right) ^2
-\frac{pd}{2a}\, \right] \tilde{s}_+=0.
\end{eqnarray}
Equation (31) contains a factor $\kappa <1$ before
$\mathcal{A}^z_i$ and formally looks like the equation of
motion of a particle moving in the reduced gauge field, which in
turn leads to an effective suppression of the Berry phase.
The calculation of Berry phase using Eq.~(31) with the gauge field
suppressed by factor $\kappa $ leads us again
to Eq.~(24).

In the absence of anisotropy and in the adiabatic approximation,
the solution of spin wave
equations has a simple form. The equations for
circular components (12) are separated
\begin{equation}
\label{32}
\pm \, i\; \frac{\partial s^{(0)}_\pm }{\partial t}
=-c_s\left( \frac{\partial }{\partial r_i}
\mp i\mathcal{A}^z_i \right) ^2 s^{(0)}_\pm \, ,
\end{equation}
and the corresponding solution is
\begin{eqnarray}
\label{33}
s_+({\bf r},t)\sim
\exp \left[ i\int k_i({\bf r})\, dr_i-i\omega t\right] ,
\end{eqnarray}
with $k_i({\bf r})\simeq g_i\omega /c_s^{1/2}+g_ig_j\, \mathcal{A}_j^z({\bf r})$.
The spin wave propagating along a closed contour $C$ acquires
the Berry phase of Eq.~(24) with $p=0$.

Using Eqs.~(7) we can present the topological field (28) in the absence
of anisotropy as
\begin{equation}
\label{34}
B_i=-\sin \theta \; \epsilon _{ijk}\;
\frac{\partial \theta }{\partial r_j}\, \frac{\partial \phi }{\partial r_k}.
\nonumber
\end{equation}
It does not depend on the angle $\psi $,
related to the choice of gauge like in the case of electromagnetism.

By creating a certain metastable configuration of the
magnetization ${\bf n}_0({\bf r})$ in the ferromagnet, we simulate
an effective gauge potential $\tilde{\mathcal{A}}_i({\bf r})$, acting on
the spin waves similar to the magnetic field in case of
electrons. In particular, when the averaged in space topological
field (28) is not zero, there arises the Landau quantization of the
energy spectrum of magnons. In the absence of anisotropy, we
find the quantized spectrum
$\hbar \omega _n=2\hbar c_s
\left| \left< {\bf B}\right> \right| (n+1/2)$, where $\left< ...\right> $
means the average in space.

\section{Interference of spin waves in mesoscopic rings}

Let us consider now the ring geometry of a ferromagnet with a
topologically nontrivial metastable magnetization ${\bf n}_0({\bf
r})$. It can be, for example, a magnetization vortex (Fig.~1a) or
an even number of domain walls in one branch of the ring like
presented in Fig.~1b. Such a magnetization profile presents a
metastable magnetic state.

\begin{figure}
\hspace*{-0.5cm}
\includegraphics[scale=0.52]{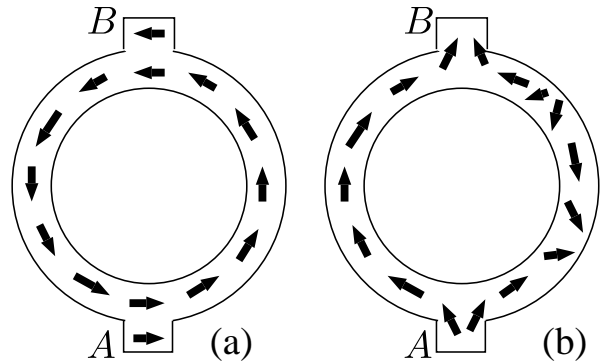}
\caption{Two rings with topologically nontrivial magnetization field.}
\end{figure}

Let us consider first the case when there is no anisotropy. If $kL\ll 1$
(adiabatic regime), the low-energy magnetic excitations of the metastable
state are described by Eq.~(32). Due to the presence of
gauge potential $\mathcal{A}^z_i({\bf r})$,
there is a phase shift of waves propagating from the
point $A$, where the waves are excited, to the observation point
$B$ (see Fig.~1).
The phase shift (Berry phase) equals to the integral $\oint
\mathcal{A}^z_i({\bf r})\, dr_i$ along the ring, and by using Stokes
theorem can be calculated as the flux $\Phi $ of topological field
$B$ defined in Eq.~(34). It can be also presented
as the spherical angle enclosed by the mapping of the ring to the
circle at the
unit sphere $S_2$. This way we can find the phase shift of $2\pi $
and $\pi N_d$ for Figs.~1a and 1b, respectively, where an even $N_d$
is the number of domain walls in the right arm of the ring.
For example, in the
case of Fig.~1b with two domain walls in the right arm, there is
no interference of spin waves excited in $A$ and coming to the
point $B$ because the corresponding phase shift is $2\pi $.

In the absence of anisotropy, the interference in the ring can
be induced by rotating all magnetic moments from the plane to
a certain angle $\alpha $ (the corresponding mapping is presented
in Fig.~2b).
The Berry phase associated with the path along the ring will be
smaller than $2\pi $. For $\alpha =\pi /6$ the Berry phase turns
out to be $\pi $. It means that the experiment with interference
of spin waves propagating from $A$ to $B$ through two different
arms of the ring, would result in a complete suppression of
the outgoing from $B$ spin wave.
Physically, it can be realized using the ring with very small
easy-plane anisotropy, $p/\omega \ll 1$ in a weak external magnetic
field along $z$ axis.

A similar idea was recently proposed by Sch\"utz {\it et al}
\cite{schutz03} for the radial orientation of magnetic moments
under inhomogeneous magnetic field directed from some point at
the axis of the ring. This magnetic field creates a "crown"
of magnetic moments, and the corresponding mapping is similar to
that presented in Fig.~2b.

\begin{figure}
\hspace*{-0.4cm}
\includegraphics[scale=0.52]{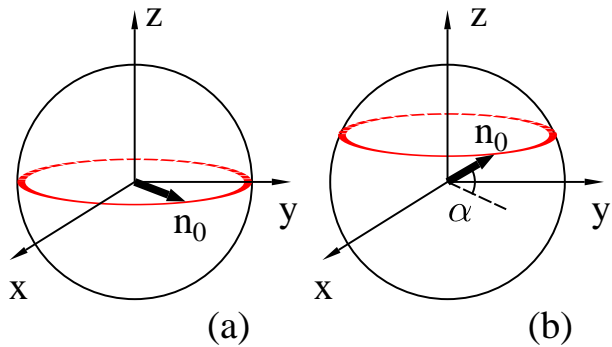}
\caption{Mapping of the ring to the ${\bf n}$-space $S_2$ (red
contour) in the case of in-plane vortex magnetization shown in
Fig.1a (a) and for the same geometry with magnetization vector
deviating from the plane to the angle $\alpha $ (b). The Berry
phase is $2\pi $ in case (a) and $\pi $ for $\alpha =\pi /6$ in
case (b).}
\end{figure}

However, in the case of nonvanishing easy-plane anisotropy, there
is no need to apply magnetic
field to provide the interference of spin waves propagating in
the geometry of Fig.~1b. In this case the Berry phase is given
by Eq.~(27) with $p=\gamma \lambda /2M_0=const$, and we obtain
the difference in phases for two waves
$\Delta \gamma _B=\pi N_d/(1+p^2/\omega ^2)^{1/2}$.
Thus, the interference of spin waves should be
clearly seen for the two-arm geometry with domain walls. The
computer simulation experiments \cite{hertel03}confirm this
expectation.

\begin{figure}
\hspace*{-0.5cm}
\includegraphics[scale=0.5]{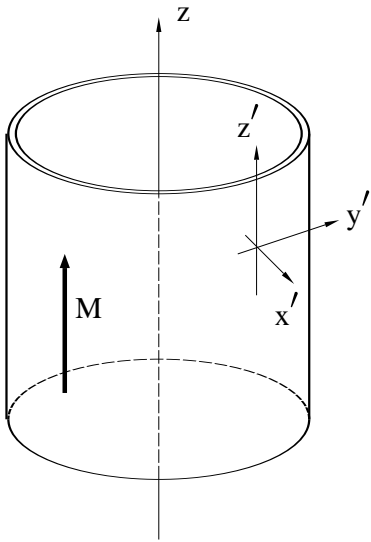}
\vspace*{-1cm}
\caption{Spin waves on a wide ring (thin-wall cylinder made of easy-plane
ferromagnetic ribbon). The magnetization is along axis $z$. }
\end{figure}

Another possibility to observe the interference of spin waves
can be presented by the geometry of a wide ring (thin-wall cylinder)
like presented in Fig.~3. Assuming the easy-plane anisotropy of
the ribbon, we obtain the ground state with a homogeneous
magnetization along the
axis of cylinder. The anisotropy axis is oriented radially in
each point of the cylinder, and the corresponding local frame
is shown in the figure as $x^\prime \, y^\prime \, z^\prime $.
Due to the homogeneous magnetization, we get $\theta =0$,
and from (7) we obtain
$\mathcal{A}^z_i({\bf r})=\partial (\psi +\phi )/\partial r_i$.
The components of anisotropy vector are $u_x=\cos \varphi $
and $u_y=\sin \varphi $. Using Eq.~(34) we find
that the condition $\tilde{u}_y=0$ reduces to $\psi +\phi =\varphi $,
and from (34) we obtain $\tilde{u}_x=1$. Correspondingly, the anisotropy
parameter $p=\gamma \lambda /2M_0$, and we obtain the Berry
phase for the closed contour on the ring
$\gamma _B(C)=2\pi N_C/(1+p^2/\omega ^2)^{1/2}$, were $N_C$ is the
winding number of the contour $C$.

The Berry phase of the spin wave propagating in magnetic ring,
plays the similar role as the phase of electron wavefunction
in the Aharonov-Bohm effect with magnetic flux penetrating
through the ring. The role of magnetic flux plays a
string through the ring.\cite{bruno04} However, the flux
created by the string does not depend on the size or shape of the
magnetic ring. Correspondingly, the Berry phase associated with
the string has the topological origin, which
makes it different from the Aharonov-Bohm effect
induced by the magnetic-field flux thorough the conductive ring.

\section{Spectrum of magnons in a ring with uniaxial anisotropy}

The other example is a ring with uniaxial anisotropy in a
homogeneous magnetic field ${\bf B}$ along the axis $z$ like presented in
Fig.~4. Due to the anisotropy and exchange interaction, the
magnetization along the ring is oriented like in Fig.~4, creating a
certain angle $\theta $ out of the ring plane.

We take the anisotropy function $\mathcal{F}$ in the form
\begin{equation}
\label{35}
\mathcal{F}\left\{ {\bf n}\right\}
=-\frac{\lambda }2\; n^2_\varphi \; ,
\end{equation}
which corresponds to the uniaxial anisotropy along the ring. The
local frame is chosen with the axis $z$ along the magnetization at
each point, and with the $z$-$x$ plane tangential to the ring
(parallel to the axis $z$). In this case we find
$\tilde{\mathcal{F}}=-\lambda s_x^2\cos ^2\theta /2$. The angles
determining the orientation of local frame are $\psi =0$ and $\phi
=r/R$.

We can calculate the angle $\theta $ using Eqs.~(1) and (32) with
vectors ${\bf u}$ and ${\bf B}$ along the axis $z$. Then, using
the polar coordinates and the condition that $n_\varphi $ and
$n_z$ do not depend on the point along the ring, we find the
energy
\begin{equation}
\label{36}
E=2\pi \zeta _0R \left( \frac{an_\varphi ^2}{2R^2}
+\frac{\lambda n_z^2}2 -M_0Bn_z\right)
\end{equation}
Substituting $n_\varphi =\sin \theta $, $n_z=\cos \theta $, we
calculate the angle $\theta $ minimizing the energy (36)
\begin{equation}
\label{37}
\cos \theta ={\rm min}\, \left\{1,\;
\frac{M_0B}{\lambda -a/R^2}\right\}
\end{equation}
for $\lambda >a/R^2$, and $\theta =0$ for $\lambda \leq a/R^2$.

\begin{figure}
\vspace*{-0.5cm}
\includegraphics[scale=0.6]{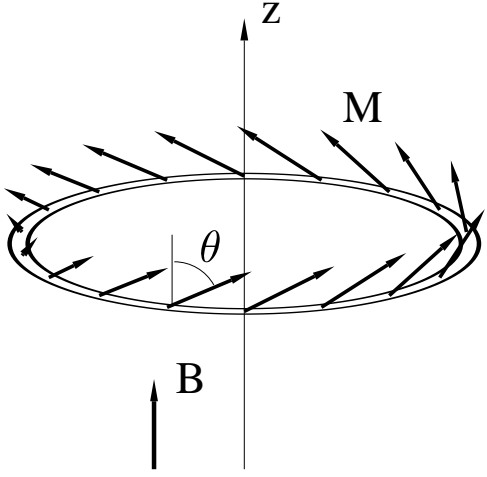}
\vspace*{-1cm}
\caption{Magnetization in a ring with uniaxial magnetic anisotropy
in a homogeneous external magnetic field along $z$ axis.}
\end{figure}

\begin{figure}
\vspace*{-1cm}
\hspace*{-1cm}
\includegraphics[scale=0.55]{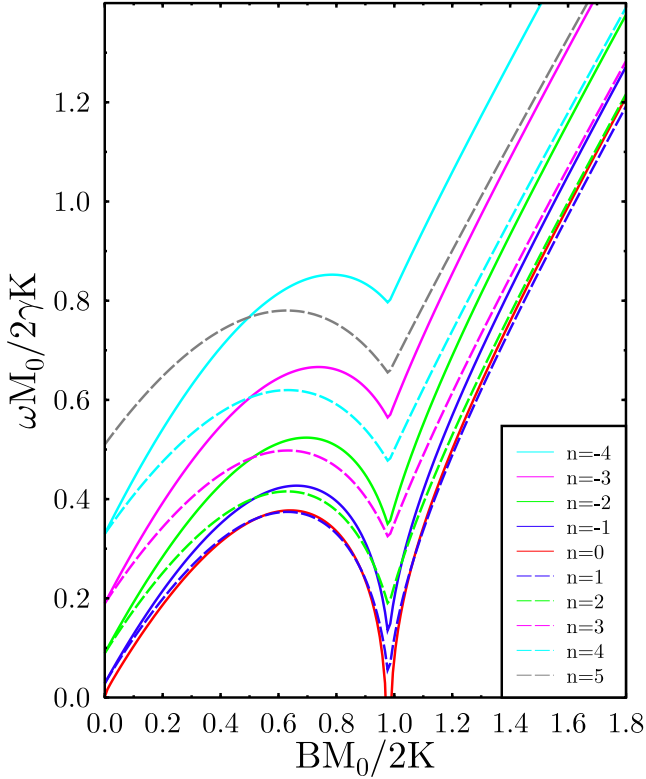}
\vspace*{-0.6cm}
\caption{Energy spectrum of magnons in the ring (Fig.~4) as a
function of external magnetic field $B$, where $K=\lambda /2$.}
\end{figure}

The gauge potential in magnetic ring induces the energy splitting
of magnons propagating in the opposite directions.\cite{bruno04}
Using Eqs.~(7) we find $\mathcal{A}^x=\sin \theta /R$,
$\mathcal{A}^y=0$, and $\mathcal{A}^z=\cos \theta $. It should be
noted that the gauge potential is constant along the ring, so that
there is no need to use the adiabatic approximation to
determine the energy spectrum of magnons.

Using Eqs.~(10) and (11) after Fourier transformation over time
$t$ and coordinate $r$ along the ring, we find
\begin{eqnarray}
\label{38}
\omega s_x=ic_s\left[
k_n^2s_y+2ik_n\mathcal{A}^zs_x+(\mathcal{A}^z)^2s_y+(\mathcal{A}^x)^2s_y\right]
\nonumber \\
+\, i\Delta _0s_y\; ,\hskip0.5cm
\nonumber \\
\omega s_y=-ic_s\left[
k_n^2s_x-2ik_n\mathcal{A}^zs_y+(\mathcal{A}^z)^2s_x\right] +ips_x
\nonumber \\
-i\, \Delta _0s_x\; ,\hskip0.5cm
\end{eqnarray}
where $\Delta _0=\gamma B$.
Here the momentum takes discrete values $k_n=2\pi n/L$ ($n=0,\pm 1,...$) to
provide the periodicity of solution for the spin wave along the ring.
We can use (38) to find the equation for elliptic components of the
spin wave
\begin{eqnarray}
\label{39}
2\left( \omega +2c_sk_n\mathcal{A}_z\right) \tilde{s}_+
=\left\{
\left[ c_sk_n^2+c_s(\mathcal{A}^z)^2+\Delta _0\right]
\frac{\eta ^2+1}{\eta }
\right. \nonumber \\ \left.
+\frac{c_s(\mathcal{A}^x)^2}{\eta }-\eta p\right\} \tilde{s}_+
+\left\{
\left[ c_sk_n^2+c_s(\mathcal{A}^z)^2+\Delta _0\right]
\frac{\eta ^2-1}{\eta }
\right. \nonumber \\ \left.
-\, \frac{c_s(\mathcal{A}^x)^2}{\eta }-\eta p\right\}
\tilde{s}_-\; ,\hskip0.5cm
\end{eqnarray}
where $\eta =d/a$ is the ellipticity.
The expression in second curved brackets vanishes for
\begin{equation}
\label{40}
\eta ^2=\frac{c_sk_n^2+c_s(\mathcal{A}^z)^2+c_s(\mathcal{A}^x)^2+\Delta _0}
{c_sk_n^2+c_s(\mathcal{A}^z)^2+\Delta _0-p}\; .
\end{equation}
Then, using (43) we find the energy spectrum of magnons in the ring
\begin{eqnarray}
\label{41}
\omega _n=\tilde{c}_s
\left( k_n-\frac{2\eta }{1+\eta ^2}\; \mathcal{A}^z\right) ^2
+\frac{c_s}2 \left( 1-\frac{4\eta }{1+\eta ^2}\right)
(\mathcal{A}^z)^2
\nonumber \\
+\, \frac{c_s}{2\eta }\; (\mathcal{A}^x)^2
-\frac{\eta p}2 +\Delta _0\frac{1+\eta ^2}{2\eta }\; .\hskip0.5cm
\end{eqnarray}
The spectrum is shown in Fig.~5 for several first values of $n$ as a function
of the magnitude of field $B$. We take the parameters: $\lambda /2=\pi M_0^2$
(it corresponds to the dipolar shape anisotropy of a magnetic cylinder),
$R=75$~nm, and $\lambda R^2/a=200$.
All the curves have a critical point $B_c$ corresponding to the magnetic
field, for which the magnetization ${\bf M}$ starts to deviate from the
direction with $\theta =0$. In view of Eq.~(37), $B_cM_0=\lambda -a/R^2$.
For $n=0$ the energy of magnons at this point is the soft mode with $\omega =0$.
This mode corresponds to a uniform rotation of spins at each point
of the ring toward a tangential to the ring direction. In the local frame
it is the uniform rotation corresponding to the state with $n=0$.

For $n\ne 0$ and $B\rightarrow 0$ we can find the spectrum in linear in $B$
approximation
\begin{eqnarray}
\label{42}
\omega _n(B)\simeq \omega _n(0)+\gamma B\left(
1+\frac1{4\left| n\right| (1+n^2)^{3/2}}\right)
\nonumber \\
-\frac{2c_snB}{\lambda R^2-a}\; ,
\end{eqnarray}
where $\omega _n(0)=\omega _{-n}(0)$.
It shows that the spectrum is degenerate at $B=0$ with respect to the sign
of $n$. The splitting for $B\ne 0$ in linear approximation gives the curve
for positive $n$ below the one for the same negative in accordance with
Fig.~5.

At large magnetic field $B$, the systematics of levels should be changed.
Namely, the lowest energy mode corresponds to the uniform {\it global}
deviation of orientation of all spins from the direction along the $z$ axis.
In the local frame, it corresponds to the mode with $n=1$. Thus, it would
be natural to label the modes with index $m=n-1$, so that the lowest in
energy is the spin wave with $m=0$.
In the limit of $B\to \infty $, each pair of modes $\pm m$ is degenerate.
It is clearly seen from Eq.~(41) with $\eta =1$. The splitting of these
modes for $B>B_c$ demonstrates the existence of {\it topological} Berry
phase for the magnons on the ring\cite{bruno04} because the equilibrium
state is the homogeneous magnetization, which leads to the vanishing
of {\it geometric} Berry phase.

\section{Conclusions}

We calculated the Berry phase associated with the propagation
of magnons in inhomogeneous ferromagnets and mesoscopic structures
with topologically nontrivial magnetization profile. We found that
the most
important effect is related to the magnetic anisotropy. Due to
the anisotropy, the Berry phase for magnons is lower than a standard
value of the spherical angle on the Berry sphere with a monopole.
Besides, an additional contribution  to the Berry phase arises in
anisotropic systems, which can be viewed as an effect of the gauge string
penetrating through the mapping contour on the unit circle.

Using these results, we demonstrated that the Berry phase can be
observable in interference experiments on magnetic rings, and
controllable using the additional homogeneous magnetic field.

\begin{acknowledgments}
V.D. thanks the University Joseph Fourier and Laboratory Louis N\'eel
(CNRS) in Grenoble for kind hospitality.
This work was supported by FCT Grant No.~POCTI/FIS/58746/2004
in Portugal, Polish State Committee for Scientific
Research under Grants Nos.~PBZ/KBN/044/P03/2001 and 2~P03B~053~25, and
also by Calouste Gulbenkian Foundation.
\end{acknowledgments}

\newpage

\end{document}